\documentclass[8.5pt,twoside,onecolumn]{article}
\usepackage[super,sort&compress,comma]{natbib} 
\usepackage{mhchem}
\usepackage{times,mathptmx}
\usepackage{balance} 
\usepackage{amsmath, amssymb} 

\usepackage{graphicx} 
\usepackage{lastpage}
\usepackage[format=plain,justification=raggedright,singlelinecheck=false,font=small,labelfont=bf,labelsep=space]{caption} 
\usepackage{fancyhdr}
\usepackage{color}
\begin{document}

\newcommand{\bea}{\begin{eqnarray}}
\newcommand{\eea}{\end{eqnarray}}
\newcommand{\kt}{k_{\rm B}T}
\newcommand{\tb}{\textcolor{blue}}
\newcommand{\tr}{\textcolor{red}}
\newcommand{\vect}[1]{\mathbf{#1}}

\makeatletter 
\def\subsubsection{\@startsection{subsubsection}{3}{10pt}{-1.25ex plus -1ex minus -.1ex}{0ex plus 0ex}{\normalsize\bf}} 
\def\paragraph{\@startsection{paragraph}{4}{10pt}{-1.25ex plus -1ex minus -.1ex}{0ex plus 0ex}{\normalsize\textit}} 
\renewcommand\@biblabel[1]{#1}            
\renewcommand\@makefntext[1]%
{\noindent\makebox[0pt][r]{\@thefnmark\,}#1}
\makeatother 
\renewcommand{\figurename}{\small{Fig.}~}

\setlength{\arrayrulewidth}{1pt}
\setlength{\columnsep}{6.5mm}
\setlength\bibsep{1pt}

\noindent\LARGE{\textbf{Strong Effect of Weak Charging in Suspensions of Anisotropic Colloids}}
\vspace{0.6cm}

\noindent\large{\textbf{Sven Dorosz,$^{\ast}$\textit{$^{a}$}, Nikhilesh Shegokar,$^{\ast}$\textit{$^{b}$}, Tanja Schilling $^{\ast}$\textit{$^{a}$}, and  Martin Oettel$^{\ast}$\textit{$^{c}$} }}\\\vspace{0.5cm}\\
{$^1$ Universit\'e du Luxembourg, Theory of Soft Condensed Matter, L-1511 Luxembourg, Luxembourg,\\$^2$ Indian Institute of Technology Bombay,
400 076 Mumbai, India,\\$^3$ Eberhard Karls Universit\"at T\"ubingen, Institut f{\"ur} Angewandte Physik, D-72076 T\"ubingen, Germany}

\vspace{0.6cm}

\noindent \normalsize
{
Suspensions of hard colloidal particles frequently serve as model systems in 
studies on fundamental aspects of phase transitions. But often colloidal 
particles that are considered as ``hard'' are in fact weakly charged. 
If the colloids are spherical, weak charging has a only a weak effect on the 
structural properties of the suspension, which can be easily corrected for. 
However, this does not hold for anisotropic particles.

We introduce a model for the interaction
potential between charged ellipsoids of revolution (spheroids) based on
the Derjaguin approximation of Debye--H\"uckel Theory
and present a computer simulation study on aspects of the system's
structural properties and phase behaviour.
In line with previous experimental observations, we find 
that even a weak surface charge has a strong
impact on the correlation functions. A likewise strong impact is seen 
on the phase behaviour, in particular,
we find stable cubatic order in suspensions of oblate ellipsoids.
}
\vspace{0.5cm}

\section*{Introduction}
Colloids are widely used as models to study basic questions of statistical 
mechanics. In particular, ``hard'' particles that only interact 
by volume exclusion have been studied intensively since the 1950s \cite{Gompper2007}. 
Hard particle systems are appealing because their phase 
behaviour is of purely entropic origin and they can easily be treated by 
computer simulation. For example, in the context of liquid crystals, 
studies of hard ellipsoids, spherocylinders and platelets have provided 
valuable insight into the basic phase transition mechanisms 
\cite{Odriozola2012, Odriozola2013, Onsager1949, Veerman1992}.

In one of the first computer simulation studies of a phase diagram of hard, 
oblate particles, Veerman and Frenkel \cite{Veerman1992} observed that 
the particles arranged parallely in stacks which in turn formed a 
suprastructure 
of perpendicular orientations. They named this phase ``cubatic''.
The existence of the cubatic phase has since been under heated debate, and 
recently several simulation studies \cite{Duncan2011, Marechal2012, Duncan2009}  showed that the cubatic is always metastable with respect to either the 
isotropic 
or the columnar phase for various round hard platelet models (i.e.~platelets with circular cross section). In contrast, the cubatic phase is stable for square plates\cite{Wilson2012}. Experimentally cubatic order has recently been 
detected in dispersions of hexagonal, charged plate-like 
particles\cite{Qazi2010}. We will show 
below that for oblate ellipsoids cubatic order becomes stabilized as 
soon as there is a small surface charge.
 
Over the past years, experimental methods to synthesize and characterize 
suspensions of ellipsoidal colloids have been advanced and theoretical 
predictions have been tested experimentally \cite{Zhang2011,Cohen2012,Kim2007,Crassous2012,Martchenko2011,Mohraz2005}. 
In 2011, Cohen et al.~\cite{Cohen2011} measured the structural properties of 
a PMMA ellipsoid system and showed significant differences when comparing 
their data to theoretical predictions for hard ellipsoids given by Percus--Yevick 
theory \cite{LetzLatz99} and simulations \cite{Talbot1990}. In the 
following, we will test our theoretical treatment of weakly 
charged ellipsoids against the results of this experimental study.
We will show that weak charging, as it is often present in PMMA-colloid 
suspensions, changes the pair correlations such that they match those observed 
experimentally and alters the phase diagram considerably.

The paper is organized as follows. We first derive the interaction potential. Then we present simulation results on the positional correlations in the system and compare them to the experimental results of ref.~\cite{Cohen2011}. 
Finally we present a scan through the phase diagram for changing surface charge density and show that the cubatic phase is stabilized.

\section{Derivation of the interaction potential}

\begin{figure}[!h]
 \includegraphics[width=\columnwidth]{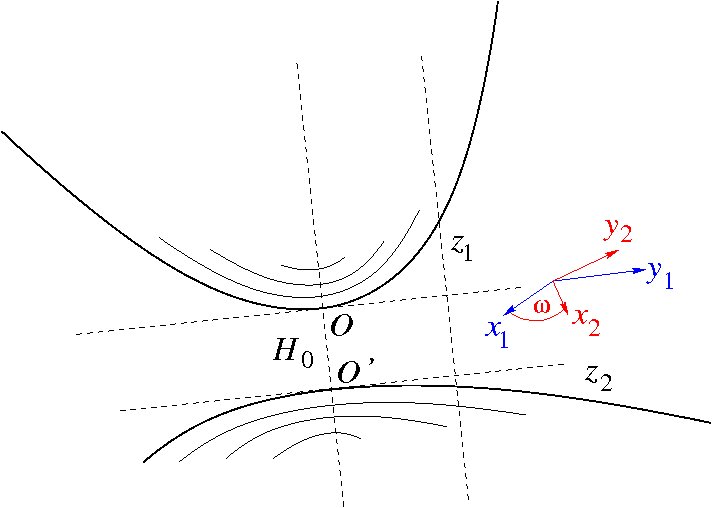}
 \caption{Two surfaces, separated by the minimal distance $H_0=\mbox{dist}(OO')$. Around the points $O$ and $O'$, the
   surfaces can be approximated by the quadratic forms $z_i = (\epsilon_i/2) x_i^2 + (\epsilon_i'/2) y_i^2$ where
   $\epsilon_i,\epsilon_i'$ are the principal curvatures of surface $i$ at point $O$ resp. $O'$ and $x_i,y_i$ are
   coordinate axes in the direction of the principal curvatures. Most generally,
   the coordinate axes $x_1$ and $x_2$ include an angle $\omega$. }
 \label{fig:derjag}
\end{figure}

The interaction of weakly charged colloids in an electrolyte suspension can be treated in an adiabatic fashion: one
assumes that co-- and counterions instantaneously readjust upon a change in the colloidal positions, giving rise 
to an effective interaction between the colloids, possibly of multi--body nature. If the Debye--H\"{u}ckel screening length
is smaller than the extensions of the colloid, then this effective potential can be well approximated by a sum of two--body terms.
For the interaction between two colloids, we use the Debye--H\"{u}ckel approximation (linearized Poisson--Boltzmann (PB) theory).
For two spheroids (ellipsoids with one rotational symmetry axis), the effective potential depends on four variables
(the center--to--center distance and three angles) such that an explicit tabulation of the PB solutions, let alone
the determination of PB solutions ``on the fly'' in a simulation code appears forbidding \cite{Chapot2004,Alvarez2010}. Here, we resort to   
 the venerable Derjaguin approximation which has been often used to calculate effective colloid--wall 
or colloid--colloid interactions
in the literature but we are not aware of its practical use in further simulation or theoretical studies
of concentrated solutions involving anisotropic particles.
The Derjaguin approximation rests upon the following argument: Suppose the free energy of the 
interaction between two planar walls is known,
and its density will be denoted by $f(h)$ where $h$ is the distance between the walls. 
The interaction potential between two convex bodies (of the same type as the walls) can be approximated 
by just integrating over geometrically opposing area elements $dA$ (at distance $h$) where the free energy of interaction between
the area elements is given by the wall free energy $f(h) dA$. The mathematical elaboration of this approximation is 
given in App.~\ref{sec:derjaguin} and results in the following expression for the
free energy of interaction $F(H_0) $ between two convex bodies with the minimal distance $H_0$ between their surfaces   
\bea
  F(H_0) &=& 
    \frac{2\pi}{\sqrt{\epsilon\epsilon'}} \int_{H_0}^\infty f(h) dh\;.
  \label{eq:fderjag}
\eea
Here, the product $\epsilon\epsilon'$ is given by 
\bea
 \label{eq:ee'}
  \epsilon\epsilon' &=& \epsilon_1 \epsilon_1' + \epsilon_2 \epsilon_2' +  \\
   & & (\epsilon_1 \epsilon_2 + \epsilon_1' \epsilon_2' ) \sin^2\omega + (\epsilon_1 \epsilon_2' + \epsilon_1' \epsilon_2 ) \cos^2\omega
 \nonumber \:.
\eea
It involves the principal curvatures $\epsilon_i$, $\epsilon_i'$ of the
surface of body $i=1,2$ in the planes tangential to the distance vector between the bodies, and also the angle $\omega$
between the coordinate systems in the tangential planes with coordinate axes given by the directions of the
principal curvatures. For these geometric definitions, see Fig.~\ref{fig:derjag}.

\subsection{Charged ellipsoids}

We will apply these ideas to the interaction between hard, charged ellipsoids (spheroids) with main axes $a$ and $b$
where $b$ is the main axis in the plane perpendicular to the rotational symmetry axis. The aspect ratio is given by $t=a/b$.
In Debye--H\"uckel approximation, the electrostatic potential $\psi$ fulfills
\bea
  \Delta \psi - \kappa^2 \psi = 0\;,
\eea
where $\kappa^{-1}$ is the Debye--H\"uckel screening length. For a charged wall with charge density $\sigma$, the solution is
\bea
  \phi_{\rm w} (z) = \phi_0 \exp(-\kappa z)
\eea
with the wall contact potential $\phi_0 = \sigma/(\epsilon_s \kappa)$ ($\epsilon_s$ is the dielectric constant of the solvent).
We approximate the solution for two charged walls at distance $h$ by
\bea
  \phi_{\rm 2w}(z) \approx \phi_{\rm w} (z)  + \phi_{\rm w} (h-z) 
\eea
and the pressure (force density per unit area between the plates) $\tilde f_{\rm 2w}$ is obtained most easily by evaluating the stress tensor at the midplane $z=h/2$
which has there only a contribution from the ion osmotic pressure:
\bea
  \tilde f_{\rm 2w} (h) = \frac{\epsilon_s}{2} \kappa^2 \phi_{\rm 2w}^2(h/2) = \frac{2\sigma^2}{\epsilon_s} \exp(-\kappa h)\;.
\eea
The free energy density $f$ is then found through integration
\bea
  f(h) & = & -\int_{\infty}^h dz \tilde f_{\rm 2w} (z) = \frac{2\sigma^2}{\epsilon_s\kappa} \exp(-\kappa h)\:.
\eea
Using this, the Derjaguin free energy (\ref{eq:fderjag}) becomes
\bea
  F &=&  \frac{2\pi}{\sqrt{\epsilon\epsilon'}} \frac{2\sigma^2}{\epsilon_s\kappa^2} \exp(-\kappa H_0)\;.
 \label{eq:fderjag_charge}
\eea

The Derjaguin free energy decays exponentially with $H_0$, as expected for the Debye--H\"uckel approximation. This decay is fast enough that
the approximation is accurate enough for practical purposes, see App.~\ref{sec:derjaguin_validity} for a discussion. Note that the anisotropy in the free energy 
has two sources: $H_0$ depends on the different orientations as well as the curvature term $1/\sqrt{\epsilon\epsilon'}$. The latter one
has a strong influence on the interaction of oblate ellipsoids (see below).
We write the Derjaguin free energy as
\bea
  \label{eq:Fderjag1}
  F(H_0) &=&  \frac{2\sigma^2 b}{\epsilon_s\kappa^2} \; V(H_0)  \;, \\
  V(H_0) &=& \frac{2\pi}{b\sqrt{\epsilon\epsilon'}} \exp(-\kappa H_0)\;.
\eea

Note that $V(H_0)$ is dimensionless in the last equation. The prefactor (with dimension of energy) 
in Eq.~(\ref{eq:Fderjag1}), $V_0=2\sigma^2 b/(\epsilon_s\kappa^2)$, contains the charge density $\sigma$
as a parameter. It is advantageous to introduce the dimensionless charge density $\tilde\sigma=\sigma e\beta/(\kappa\epsilon_s)$
(where $e$ is the elementary charge and $\beta = 1/(k_{\rm B} T)$ is the inverse temperature) as well as the Bjerrum length of the solvent, 
$\lambda_B= \beta e^2/(4\pi\epsilon_s)$. With these definitions the prefactor becomes 
\bea \label{eq11}
  \beta V_0 = \frac{\tilde\sigma^2}{2\pi} \frac{b}{\lambda_B} \;.
\eea
Below, in the comparison with the experiments of ref.~\cite{Cohen2011}, we will treat $\tilde\sigma$ as a fitting parameter. Regarding the
interpretation of this value one should keep in mind that it is an effective or renormalized charge density. For $\tilde \sigma \lesssim 1$,
the bare and the renormalized charge density are approximately the same, whereas in the limit of very large bare charge densities
the renormalized charge density approaches a constant, $\tilde \sigma \to 4$ \cite{Bocquet2002}. Thus the approach is consistent only
for $\tilde \sigma < 4$.

\subsection{Numerical implementation}
\label{sec:num}

We computed numerically the dimensionless, exponentiated free energy $\exp(- V(H_0))$ (Eq.~\ref{eq:Fderjag1})
on a 4--dimensional grid with axes characterizing the relative configurational state of two ellipsoids  
(center--to--center distance and three angles).
For each configuration, the minimal distance $H_0$ was determined by a conjugate gradient routine and the
radii and directions of principal curvature were determined through the first and second fundamental form
of the ellipsoid surfaces at the two points $O$ and $O'$ (whose distance is $H_0$, see Fig.~\ref{fig:derjag}).
In the Monte Carlo simulations (see below), the such tabulated free energy was used as the acting potential
between pairs of ellipsoids, together with linear interpolation to
determine the potential at off--grid values of the variables characterizing the relative configurational state.

We also tested a further approximation to the Derjaguin free energy in which the curvature term $1/\sqrt{\epsilon\epsilon'}$
is replaced by a constant, the average radius of the ellipsoid. Then $V(H_0)$ only depends on the minimal distance $H_0$, which can be
well approximated by an extension of the Perram--Wertheim routine\cite{Per85} frequently used for checking overlap of hard ellipsoids
(see App.~\ref{sec:pw}). In this way, the potential can be determined ``on the fly'', and it works reasonably well for aspect ratios 
$0.8 \lesssim t \lesssim 2$. Note, however, that the short--range anisotropy of the potential increases rapidly with the aspect ratio becoming small. 
For oblate ellipsoids ($t<1$, disk--like particles),
the ratio between the potential at contact in side--side configuration (flat sides of the disk touching) and in edge--edge configuration
(rims of the disk are touching) is $(t^2+1)/(2t^3)$  and thus scales for small $t$ as $1/t^3$. Therefore, this further approximation
to the Derjaguin free energy is not applicable to flat oblates.

\section{Simulation Results}
\label{sec:g}

We have carried out Monte Carlo simulations at constant temperature T, constant number of particles N and volume V with periodic boundary conditions, and computed equilibrium structural properties of the system as a function of the effective surface charge density and the packing fraction. The particle number $N$ ranged from 3000 to 3840, T was set to 300K.

First, we discuss the structure of the isotropic phase. This part of our work 
has been inspired by recent experimental measurements  
of the radial, orientation--averaged pair correlation function $g(r)$ in  
suspensions of prolate ellipsoids (aspect ratio $t=1.6$, $a=3.2$ $\mu$m,  $b=2.0$ $\mu$m) and oblate ellipsoids ($t \approx 0.25$, $a\approx0.96$ $\mu$m $b\approx 3.8$ $\mu$m, with a considerable experimental uncertainty on the polydispersity and thus on the value of $t$)\cite{Cohen2011}. 

The structural correlations that are presented in 
ref.~\cite{Cohen2011} are much stronger than one would expect for a system of 
hard ellipsoids, and this was taken as an indication
that on the theory side, the correlations in suspensions of ellipsoids are 
not sufficiently well understood.
However, the experimental suspension was additionally stabilized by a surfactant which introduced a small amount of charge on the particles.
In a later study \cite{Cohen2012}, the authors investigated the influence of charge on $g(r)$ for the prolate particles by simulation and 
found it non--negligible: with a small charge density of $\sigma \approx 9$ $e/\mu$m$^2$ (where $e=1.6\cdot10^{-19}C$ is the elementary charge), distributed on particles modelled by an assembly of three
cut spheres to approximate the shape of the ellipsoids, the experimental $g(r)$ could be reproduced. No corresponding results for the oblate particles
have been reported, though.   

\begin{figure}
\centering
\includegraphics[width=\columnwidth]{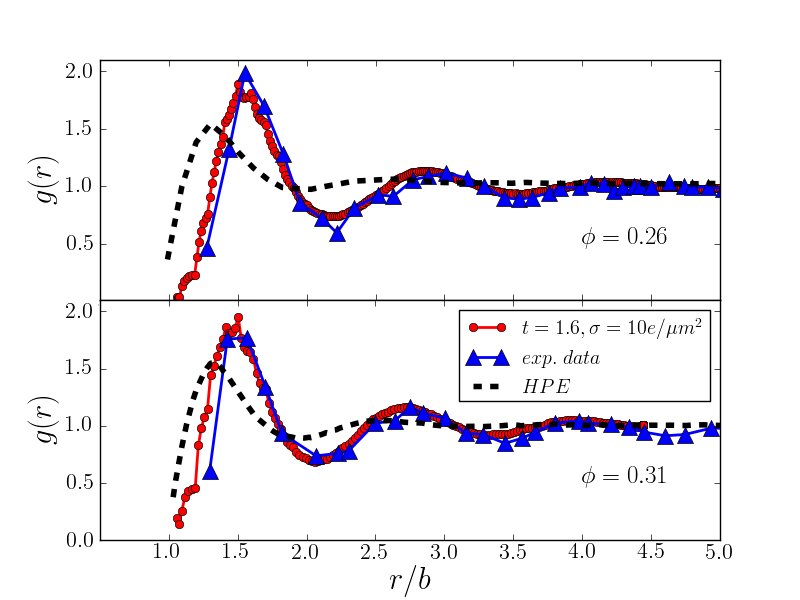}
\caption{Radial positional distribution function of ellipsoids with aspect ratio $t=1.6$ and packing fractions $\phi=0.26$ (top) and $\phi=0.31$ (bottom), simulation data (circles), experimental data (triangles) and data for hard prolate ellipsoids (HPE, dashed line) taken from Fig.~2(a) and 2(b) in ref.~\cite{Cohen2011}.}
\label{fig:gr1}
\end{figure}

In order to model the experiment of ref.~\cite{Cohen2011},
we set $\lambda_B\approx 22$ nm appropriate for a solvent with an average dielectric constant 
of $\epsilon_s=2.5$ and $\kappa^{-1} = 0.3$ $\mu$m for the Debye--H\"uckel screening length.
We only vary the effective surface charge density 
to reproduce the experimental data. 
Fig.~\ref{fig:gr1}(top) shows the radial positional distribution function $g(r)$ of ellipsoids (circles) for an aspect ratio $t=1.6$ and a packing fraction $\phi=0.31$. 
The simulation data perfectly match the experimental data (triangles, from Fig.~2 in ref.~\cite{Cohen2011}). 
The corresponding dimensionless charge density is given by $\tilde\sigma = 0.83$, i.~e.~the effective charge density is 
$\sigma =10$ $e/\mu$m$^2$. This value is reasonable for the experimental system used in ref.~\cite{Cohen2011}, it is in good agreement with the deduced effective charge on the colloids of ref.~\cite{Cohen2012}, and it justifies in retrospect the assumption of the Debye--H\"uckel approximation used to derive the interaction potential. (Note that the modelling of the electrostatic 
particle interactions in ref.~\cite{Cohen2012} approximately corresponds to the Derjaguin free energy with curvature neglected (see Sec.~\ref{sec:num} above) which works well for
the moderate aspect ratio of 1.6 but not for particles with larger curvatures. This rationalizes the good agreement between our results and those of ref.~\cite{Cohen2012} for $\sigma$ and $g(r)$.)     

The second data set discussed in ref.~\cite{Cohen2011} has been measured in a more dilute suspension of prolate ellipsoids at a packing fraction $\phi=0.26$. Using the same system parameters for 
$\lambda_B$, $\kappa^{-1}$ and even $\sigma$ as for $\phi=0.31$, we again obtain very good agreement of the radial distribution functions, see Fig.~\ref{fig:gr1}(bottom). The data set for hard ellipsoids from ref.~\cite{Cohen2011} is presented as well for comparison. 

\begin{figure}
 \centering
\includegraphics[width=\columnwidth]{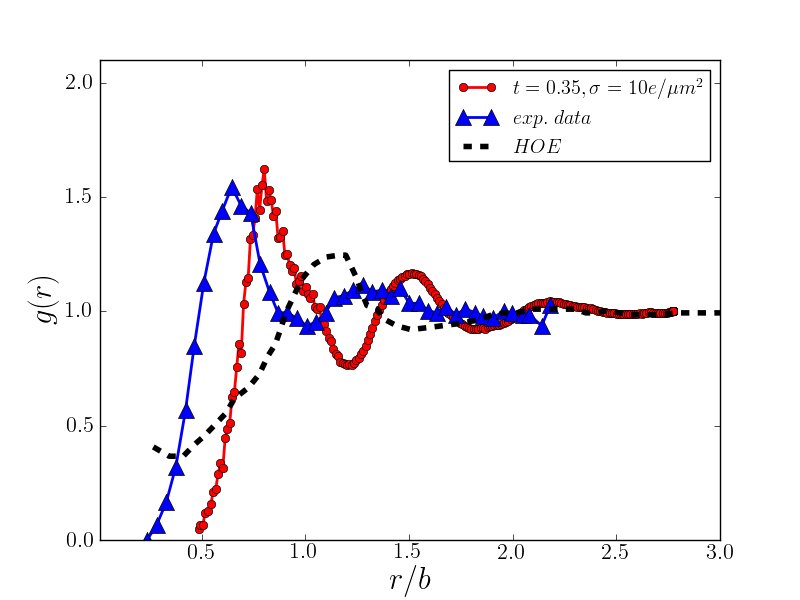}
 \caption{Radial positional distribution function of oblate ellipsoids 
with a packing  
fraction of $\phi=0.35$, simulation data (circles) for $t=0.35$, 
experimental data for ''$t\approx0.25$'' 
(triangles) and data for hard oblate ellipsoids (HOE) taken from 
Fig.~3 in ref.~\cite{Cohen2011} (dashed line, data on the abscissa is 
multiplied 
by a factor $1.4$ with respect to ref.~\cite{Cohen2011} to undo the rescaling 
and recover units of $b$.).
}
 \label{fig:gr3}
\end{figure}

The last radial distribution function that is presented in 
ref.~\cite{Cohen2011}, was measured in a suspension of oblate ellipsoids 
of an aspect ratio of ``$t\approx0.25$'' (with a larger polydispersity 
than in the prolate case). As explained in Sec.~\ref{sec:num},
the curvature around the rim of the particles in this case is important for the electrostatic interactions, hence the approach of ref.~\cite{Cohen2012} could not be applied here. 
Fig.~\ref{fig:gr3} shows our simulation results for oblate ellipsoids. We set again the same value for $\lambda_B$, $\kappa^{-1}$ and $\sigma$. 
The experimental and theoretical data agree reasonably well
given the uncertainty of the aspect ratio of the experimental system.

To conclude this section, we validated the Derjaguin approximation for the electrostatic interaction of charged ellipsoids. A small amount of surface charge has a strong
influence on the pair correlations, due to the small dielectric constant (large Bjerrum length) of the solvent. The short--range anisotropy of the electrostatic interaction
is especially important for oblate ellipsoids.

\section{Impact of the surface charges on the nematic phase}

We now consider oblate ellipsoids of aspect ratio $t=0.25$ at a packing 
fraction of $\phi=0.48$. In suspensions of hard ellipsoids  
the nematic phase is located at packing fractions 
$\phi> 0.4$ \cite{Odriozola2012}. 
We study the effect of increasing surface charge density, ranging from 
$\sigma=0.00002$ $e/\mu$m$^2$ to $\sigma=2.0$ $e/\mu$m$^2$, on the structural 
properties of the liquid. 
Note that these surface charge densities are even lower than the value that was needed to reproduce the experimental findings of ref.~\cite{Cohen2011}. 
The Bjerrum length $\lambda_B$ and the Debye--H\"uckel screening length $\kappa^{-1}$ are not modified with respect to the previous section.

\begin{figure}
\includegraphics[width=\columnwidth]{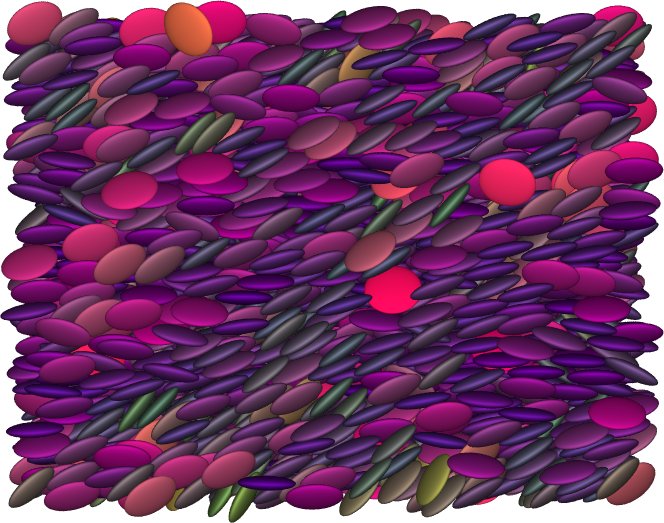}
\caption{Snapshot of the suspension of almost hard oblates ($\sigma=0.00002$ $e/\mu$m$^2$) (aspect ratio $t=0.25$) at a packing fraction $\phi = 0.48$ in the nematic phase. Colour code according to orientation.
}
\label{fig:gr5}
\end{figure}

\begin{figure}
\includegraphics[width=\columnwidth]{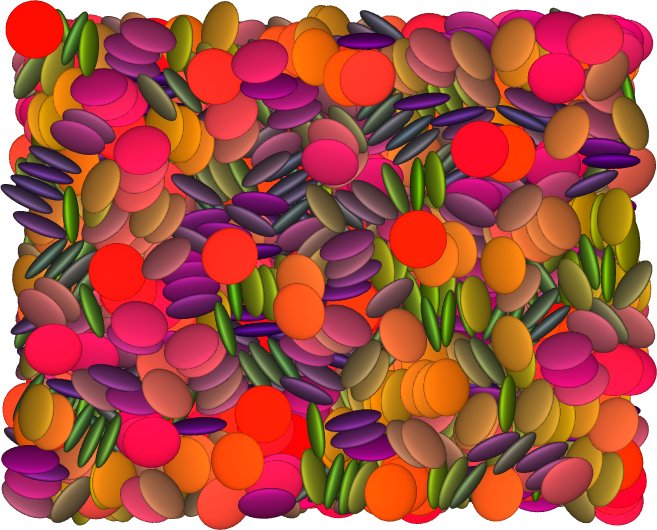}
\caption{Snapshot of the suspension of charged oblates at surface charge density $\sigma=0.2$ $e/\mu$m$^2$ (aspect ratio $t=0.25$) at a packing fraction $\phi= 0.48$. Colour code according to orientation.
}
\label{fig:gr6}
\end{figure}

\begin{figure}
\includegraphics[width=\columnwidth]{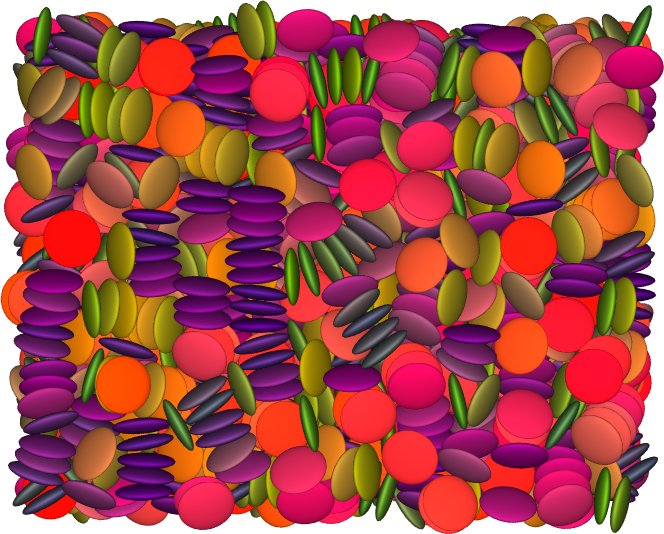}
\caption{Snapshot of the suspension of charged oblates at surface charge density $\sigma=2.0$ $e/\mu$m$^2$ (aspect ratio $t=0.25$) at a packing fraction $\phi= 0.48$. Colour code according to orientation.
}
\label{fig:gr7}
\end{figure}

As the initial configuration of a first set of simulation runs we used an fcc crystal in which the ellipsoids were oriented in parallel. We let the system relax until its energy had reached a stable value. In the case of perfect charge screening (i.e.~almost hard ellipsoids) the system relaxed into the expected nematic phase, see Fig.~\ref{fig:gr5}.
A similar degree of nematic ordering formed for a surface charge density of $\sigma=0.02$ $e/\mu$m$^2$. In contrast
we observe qualitatively different behaviour for surface charge densities $\sigma=0.2$ $e/\mu$m$^2$ and $\sigma=2.0$ $e/\mu$m$^2$, see snapshots in Fig.~\ref{fig:gr6} and Fig.~\ref{fig:gr7}. Note that these configurations evolved from an initial fcc configuration with parallel orientation of the ellipsoids.

\begin{figure}
\includegraphics[width=\columnwidth]{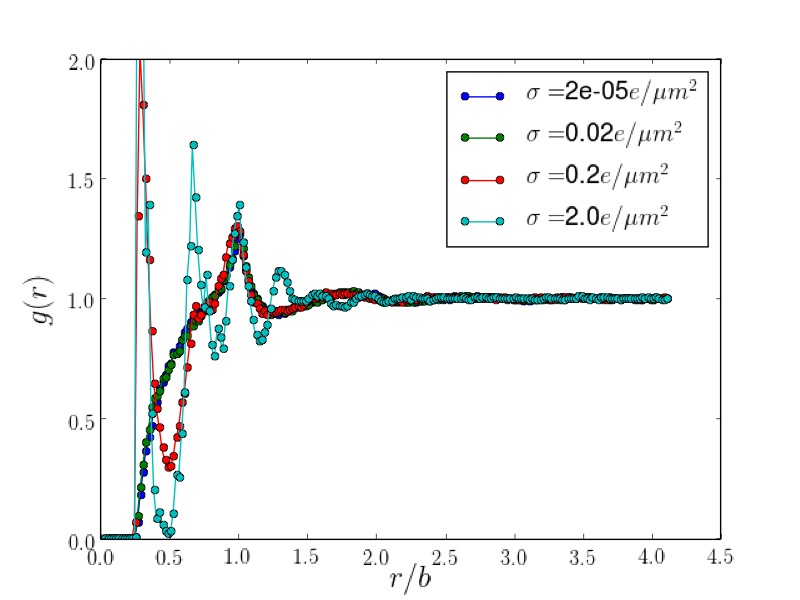}
 \caption{Radial positional distribution, aspect ratio $t=0.25$, for different surface charge densities $\sigma$ at a packing fraction $\phi=0.48$.}
 \label{fig:gr8}
\end{figure}

Fig.~\ref{fig:gr8} shows $g(r)$ for different surface charge densities $\sigma$ at $\phi=0.48$. 
In the nematic phase, $g(r)$ has a cusp at a distance $r$ that corresponds to in-plane rim--rim configurations. In contrast, at $\sigma = 0.2$ $e/\mu$m$^2$ there is pronounced positional order with peak positions at multiples of the length of the small axis $a$ plus a small distance to account for the electrostatic repulsion.

\begin{figure}
\includegraphics[width=\columnwidth]{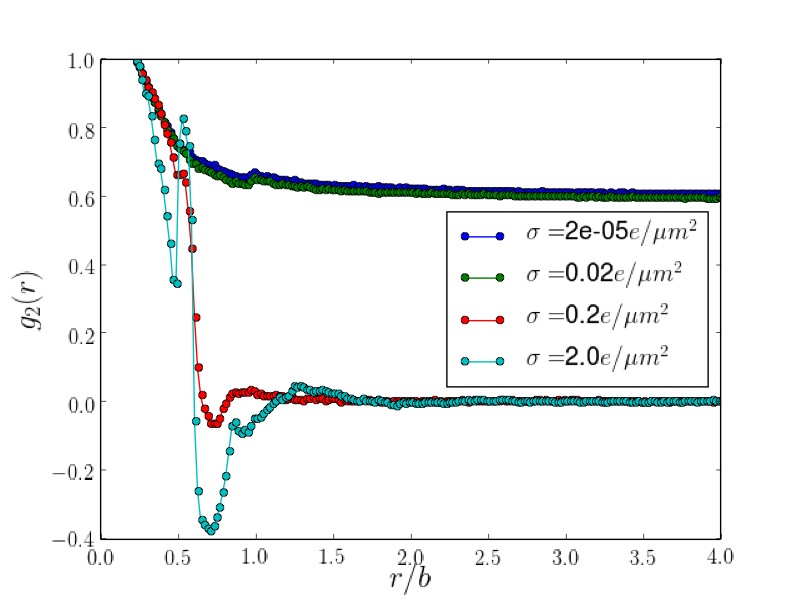}
 \caption{Radial orientational distribution, aspect ratio $t=0.25$, for different surface charge densities $\sigma$ at a packing fraction $\phi=0.48$.}
 \label{fig:gr9}
\end{figure}

Fig.~\ref{fig:gr9} shows the orientational distribution function 
\[
g_2(r) = \frac{1}{g(r)} \frac{1}{2} \langle 3 {\bf u}_i{\bf u}_j - 1\rangle \quad ,
\]
where ${\bf u}_i$ is the unit vector along the axis of particle $i$ and the 
average is over all pairs of particles and the canonical ensemble.
At small $\sigma$ $g_2(r)$ decays smoothly to a non-vanishing value at large distances, which is characteristic for nematic ordering. 
At $\sigma = 0.2$ $e/\mu$m$^2$ and above there is parallel order at short distances and random orientation at large distances, i.e $g_2(r)$ decays to zero. The first ``perpendicular peak'' with $g_2 < 0$ appears at the distance that corresponds to a configuration in which the rim of one ellipsoid points to the pole of the other. This peak is superposed with the second layer of parallel stacking in $g(r)$. We conclude that stacks of ellipsoids are arranged perpendicular to each other to form cubatic order.

To test whether this phase is metastable, we then initialized 
simulations in the nematic phase, the columnar phase and a perfect long-range 
cubatic phase. All runs equilibrated into the phase discussed above, thus we 
conclude that it is the most stable.

\section{Conclusion}

We have added surface charges to hard ellipsoids and treated them numerically using the Derjaguin approximation. With the Derjaguin approximation, the short--range anisotropy of the electrostatic interactions is captured
quantitatively correctly as long as the screening length is smaller
than the extensions of the particles. Even very small surface charges, as they often are present in experiments on ``hard'' particles have a strong effect on the structure of the suspension.

We showed for charged oblate ellipsoids that a phase of perpendicularly oriented short stacks (a cubatic) is thermodynamically more stable than the 
nematic phase, which is stable for uncharged ellipsoids at the same packing fraction. The cubatic phase is also more stable than the crystal and the columnar phase. It should be accessible to experiments on suspensions of PMMA ellipsoids.

\section*{Acknowledgements}

This project has been financially supported by the DFG (SFB Tr6 and SPP1296) and by the National Research Fund,
 Luxembourg co-funded under the Marie Curie Actions of the European Commission (FP7-COFUND) and under the project FRPTECD. Computer simulations presented 
in this paper were carried out using the HPC facility of the University of Luxembourg.

\begin{appendix}

\section{Derjaguin approximation: mathematical derivation}
\label{sec:derjaguin}

The Derjaguin approximation \cite{Der34} 
involves the following steps: Suppose the free energy of the interaction between two planar walls is known, 
and its density will be denoted by $f(h)$ where $h$ is the distance between the walls. For calculating the interaction potential between two convex bodies (of the same type as the walls) one determines
the minimal distance $H_0$ between the two surfaces and the tangential planes (see Fig.~\ref{fig:derjag}). 
The free energy of interaction between the two bodies 
is approximated by
\bea
  F &=& \int \int dx dy f(H_0 + z_1 +z_2), 
\eea  
where the integral runs over (one of) the tangential planes and $z_1,z_2$ are the distances of the point on surface $i$ described by $z_i(x,y)$ to their respective tangential plane
(see Fig.~\ref{fig:derjag}). If $f(h)$ quickly decays to zero, it is safe to integrate over the whole plane.  

This integral is greatly simplified if we approximate the surfaces by quadratic forms around the points $O$ and $O'$
respectively:
\bea
  z_1 & =& \frac{\epsilon_1}{2} x_1^2 + \frac{\epsilon_1'}{2} y_1^2 \\
  z_2 &=& \frac{\epsilon_2}{2} x_2^2 + \frac{\epsilon_2'}{2} y_2^2  \:,
\eea
where $\epsilon_1, \epsilon_1'$ are the principal curvatures of surface 1 at point $O$ and $x_1,y_1$ are coordinates in the
tangential plane in the direction of the principal curvatures. Likewise for surface 2. The directions of the principal curvatures
of surface 1 and 2 do not agree but include an angle $\omega$, thus:
\bea
  \begin{pmatrix} x_2 \\ y_2 \end{pmatrix} &=&
  \begin{pmatrix} \cos\omega & \sin\omega \\ -\sin\omega & \cos\omega \end{pmatrix}
  \begin{pmatrix} x_1 \\ y_1 \end{pmatrix} \:.
\eea
The distance $z_1+z_2$ becomes
\bea
  \label{eq:z_1}
  z_1 + z_2 &  = & \frac{1}{2} \begin{pmatrix} x_1 & y_1 \end{pmatrix}
                \begin{pmatrix} A & C \\ C & B \end{pmatrix}
                  \begin{pmatrix} x_1 \\ y_1 \end{pmatrix}  \qquad {\rm with} \\
     & & A = \epsilon_1 + \epsilon_2 \cos^2\omega + \epsilon_2' \sin^2\omega \\
     & & B = \epsilon_1' + \epsilon_2' \cos^2\omega + \epsilon_2 \sin^2\omega  \\
     & & C = (\epsilon_2 - \epsilon_2') \cos\omega\sin\omega  \; .
\eea
This distance is a quadratic form. We may perform a rotation to another coordinate system $x,y$ where the off--diagonal matrix elements
become zero, i.e.
\bea
   z_1  + z_2 &  = & \frac{1}{2}  \begin{pmatrix} x & y \end{pmatrix}
                \begin{pmatrix} \epsilon & 0 \\ 0 & \epsilon' \end{pmatrix}
                  \begin{pmatrix} x \\ y \end{pmatrix} \; .
  \label{eq:z_2}
\eea  
Using that result, the free energy becomes
\bea
  F &=& \int \int dx dy f(H_0 + \epsilon x^2/2 + \epsilon' y^2/2 )\:. 
\eea  
Introducing new coordinates $r,\phi$ via $x=r\cos\phi/\sqrt{\epsilon}$ and $y=r\sin\phi/\sqrt{\epsilon'}$:
\bea
  F &=& \int_0^\infty r dr \int_0^{2\pi} d\phi  f(H_0 + r^2/2 ) \\
   &=& \frac{2\pi}{\sqrt{\epsilon\epsilon'}} \int_{H_0}^\infty f(h) dh
  \label{eq:fderjag1}
\eea
where the second line (which is Eq.~(\ref{eq:fderjag}))
follows from the substitution $h=H_0+r^2/2$.
The force (in direction of $OO'$) between the two bodies is just given by
\bea
  K &=& - \frac{\partial F}{\partial H_0} = \frac{2\pi}{\sqrt{\epsilon\epsilon'}} f(H_0) \:.
\eea
The product  $\epsilon\epsilon'$ is the determinant of the matrix in Eq.~(\ref{eq:z_2}) and must be equal to the
determinant of the matrix in Eq.~(\ref{eq:z_1}). Thus we find:
\bea
  \epsilon\epsilon' &=& \epsilon_1 \epsilon_1' + \epsilon_2 \epsilon_2' +  \\
   & & (\epsilon_1 \epsilon_2 + \epsilon_1' \epsilon_2' ) \sin^2\omega + (\epsilon_1 \epsilon_2' + \epsilon_1' \epsilon_2 ) \cos^2\omega
 \nonumber 
\eea
which is Eq.~(\ref{eq:ee'}).

\section{Validity of the Derjaguin approximation}
\label{sec:derjaguin_validity}

We can estimate the validity of the Derjaguin approximation by considering the example of two interacting, charged spheres 
with radius $r_0$ and charge density $\sigma$ at center distance $d$ for which we can
compare the ``exact'' Debye--H\"uckel result with the corresponding Derjaguin approximated result. 
In Debye--H\"uckel approximation, the potential of a single sphere is given by
\bea
  \Phi_s (r) &=& \frac{Q_{\rm eff}}{4\pi\epsilon_s} \frac{\exp(-\kappa r)}{r}\;,
\eea
and the charge $Q_{\rm eff}$ is determined through the boundary condition $\left.\partial\Phi/\partial r\right|_{r=r_0} = - \sigma/\epsilon_s$, giving
\bea
  Q_{\rm eff} &=& \frac{4 \pi r_0^2 \exp(\kappa r_0)}{1+\kappa r_0} \; \sigma \;. 
\eea
In superposition approximation (as before), the interaction free energy of two such spheres is given by
\bea
  F_s &=& \frac{Q^2_{\rm eff}}{4\pi\epsilon_s} \frac{\exp(-\kappa d)}{d} \nonumber \\
     &=& \frac{4\pi r_0^4}{(1+\kappa r_0)^2} \frac{\sigma^2}{\epsilon_s} \frac{\exp(-\kappa H_0)}{H_0+2r_0} \;,
  \label{eq:fs}
\eea 
where $H_0 = d-2r_0$ is the minimal surface--to--surface distance. The Derjaguin approximated free energy follows from
Eq.~(\ref{eq:fderjag_charge}) using $\epsilon\epsilon' = 4/r_0^2$:
\bea
  F_D &=& \frac{2\pi r_0}{\kappa^2} \frac{\sigma^2}{\epsilon_s} {\exp(-\kappa H_0)}\;.
\eea
It is the limit $\kappa r_0 \gg 1, H_0 \ll r_0$ of Eq.~(\ref{eq:fs}). The ratio is given by
\bea
  \frac{F_D}{F_s} &=& \left(  1+ \frac{H_0}{2r_0}  \right) \left( 1 + \frac{1}{\kappa r_0}\right)^2 \;.
\eea
From this equation one sees that the Derjaguin approximation produces a large relative error for $H_0 \gtrsim 2r_0$.
Appropriate for the examples studied in Sec.~\ref{sec:g}, we set $r_0=1$ $\mu$m and the screening length $\kappa^{-1}= 300$ nm.
Thus, $\kappa r_0 \approx 1/3$,
and the Derjaguin approximation overestimates the free energy by a factor 3.4 at $H_0=2r_0=2$ $\mu$m.
However, at that distance the potential has dropped by a factor $\exp(-2\kappa r_0) \approx 0.0013 $ compared to its value at contact,
so the free energy itself at that distance is small and likewise the absolute error the Derjaguin approximation produces is small.
The effect is presumably negligible in our simulations which are sensitive to the short--range behaviour of the effective free energy between the particles.

\section{Perram--Wertheim approximation for the minimal distance}
\label{sec:pw}

Perram and Wertheim \cite{Per85} developed a criterion to check for overlap of two ellipsoids. The algorithm can also be used to
compute an approximate minimal distance. We summarize this approach as follows:
Suppose we have two ellipsoids with half axes $a_1, a_2, a_3$ and corresponding axis orientation vectors (of unit length)
in a lab--fixed coordinate system $\vect u_1, \vect u_2, \vect u_3$
(ellipsoid 1) and $\vect v_1, \vect v_2, \vect v_3$ (ellipsoid 2). One defines two matrices:
\bea
  A &=& \sum_{k=1}^3 a_k^{-2} \vect u_k \vect u_k^T \;, \\
  B &=& \sum_{k=1}^3 a_k^{-2} \vect v_k \vect v_k^T \;.
\eea    
Let $\vect r_a$ and $\vect r_b$ denote the center positions of ellipsoid 1 and 2, respectively. We define quadratic forms
\bea
   F_A &=& (\vect r - \vect r_a)^T A (\vect r - \vect r_a) \;, \\
   F_B &=&  (\vect r - \vect r_b)^T B (\vect r - \vect r_b) \;,
\eea 
and the ellipsoid surfaces are given by the solutions to the equations $F_A =1$ and $F_B=1$. For points inside the ellipsoid,
$F_{A[B]}<1$, for points outside $F_{A[B]}>1$.  
Further we define a quadratic form
\bea
  F(\vect r, \lambda) = \lambda F_A + (1-\lambda) F_B \;.
\eea
and its minimum, depending on $\lambda \in [0,1]$:
\bea
  F(\vect r (\lambda), \lambda) = {\rm min}_{\vect r} F(\vect r, \lambda) \;.
 \label{eq:nablaF}
\eea
For $\lambda=0$,  $\vect r(0)=\vect r_b$ and for $\lambda=1$,  $\vect r(1) = \vect r_a$. Thus as $\lambda$ varies from 0 to 1,
then $\vect r (\lambda)$ moves from the center of ellipsoid 2 to the center of ellipsoid 1. If the two ellipsoids do not overlap,
then there exists a particular $\lambda$ for sure for which  $\vect r (\lambda)$ is outside both ellipsoids and 
$ F(\vect r (\lambda), \lambda)>1$ there. If the ellipsoids overlap, then $ F(\vect r, \lambda) < 1$ in the overlap region  and
thus for each $\lambda$ the minimal point $\vect r(\lambda)$  can not lie outside both ellipsoids since there $F(\vect r, \lambda) >1$.
Therefore a useful overlap criterion is formulated with introducing
\bea
   s = {\rm max}_{\lambda\in(0,1)} F(\vect r (\lambda), \lambda)
 \label{eq:sdef}
\eea
which fulfills
\bea
  s  \left\{ \begin{matrix} >1 & \qquad {\rm overlap} \\ =1 & \qquad {\rm tangent} \\ <1 & \qquad {\rm no\; overlap} \end{matrix} \right. \;.
\eea 
This criterion is convenient to use due to the explicit form for $F(\vect r (\lambda), \lambda)$ which Perram and Wertheim provide
\cite{Per85}:
\bea
  F(\vect r (\lambda), \lambda) &=& \lambda(1-\lambda) (\vect r_b - \vect r_a)^T C (\vect r_b - \vect r_a) \;, \\
                             C &=& (\lambda B^{-1} + (1-\lambda) A^{-1})^{-1} \;. 
\eea
The maximization needed in Eq.~(\ref{eq:sdef}) has to be done numerically, though. 

The value of $s$ can also be used to calculate an approximative minimal distance through
\bea
  d_{\rm PW} &=& |\vect r_b - \vect r_a| \left( 1 - \frac{1}{\sqrt{s}} \right) \;.
 \label{eq:dpw}
\eea
The interpretation, according to Paramonov and Yaliraki \cite{Par05}, is as follows. First, the geometrical meaning of the
$\lambda$ maximization in Eq.~(\ref{eq:sdef}) becomes clear by looking at
\bea
   \frac{d F(\vect r(\lambda), \lambda)}{d\lambda} &=& F_A(\vect r(\lambda), \lambda) -  F_B(\vect r(\lambda), \lambda) \nonumber\\
						   & & +\frac{d\vect r}{d\lambda} \cdot \nabla F(\vect r(\lambda), \lambda) \;.
\eea
The term $\nabla F$ is zero by virtue of the definition of $ F(\vect r(\lambda), \lambda)$ in Eq.~(\ref{eq:nablaF}) and thus
the derivative above is zero when $s=F_A(\vect r(\lambda_{\rm max}),\lambda_{\rm max})=F_B(\vect r(\lambda_{\rm max}), \lambda_{\rm max})$. This, however,
describes the tangential contact between ellipsoids with {\em scaled} half--axes $\sqrt{s} a_1, \sqrt{s} a_2,\sqrt{s} a_3$.
($\vect r(\lambda_{\rm max})$ is the tangential contact point since $\nabla F_A  || \nabla F_B$ there.) 
Then the points $\vect s_a, \vect s_b$ defined by
\bea
   \vect s_a - \vect r_a = \frac{1}{\sqrt{s}}(\vect r(\lambda_{\rm max}) - \vect r_a)
\eea
(likewise for $a\to b$) lie on the surface of ellipsoid 1 and 2, respectively. We see that $\vect s_b - \vect s_a$ is 
parallel to the center distance vector $\vect r_b - \vect r_a$ and that $d_{\rm PW} = |\vect s_b - \vect s_a|$. Thus 
$d_{\rm PW}$ is the minimal {\em directional} distance between the ellipsoids (i.e. minimal distance between 
two points on the surfaces of ellipsoid 1 and 2 in the direction of the center distance vector.

\begin{figure}
 \begin{center}
   \includegraphics[width=\columnwidth]{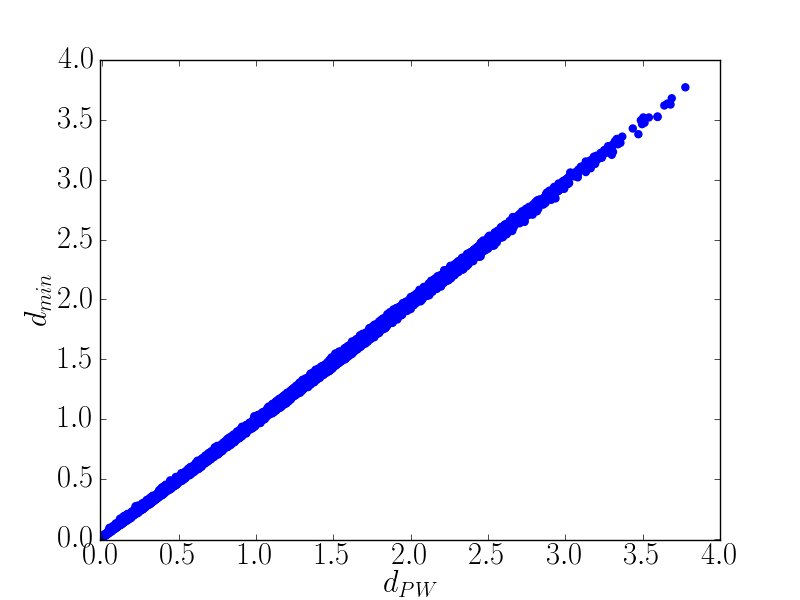}
 \end{center}
 \caption{ For given configurations of two prolates of aspect ratio $t=1.6$, we plot the minimal distance relative to the Perram Wertheim distance.}
 \label{fig:v}
\end{figure}

This approximation is not exact but as presented in Fig. \ref{fig:v} the error is small and hence the approach using the simple determination of the Perram--Wertheim distance justified for the calculation.

The relation is not one to one because of the different relative orientations but there is agreement for moderate aspect ratios.

\end{appendix}


\providecommand*{\mcitethebibliography}{\thebibliography}
\csname @ifundefined\endcsname{endmcitethebibliography}
{\let\endmcitethebibliography\endthebibliography}{}

\end{document}